\documentclass[prb,twocolumn,amssymb,showpacs]{revtex4}

\usepackage{bm}% bold math
\usepackage{graphicx}% Include figure files
\usepackage{dcolumn}% Align table columns on decimal point
\begin{document}

\title{A multiexcitonic quantum dot in an optical microcavity}
\author{Herbert Vinck, Boris A. Rodriguez}
\affiliation{Instituto de Fisica, Universidad de Antioquia, AA 1226, Medellin, Colombia}

\author{Augusto Gonzalez}
\affiliation{Instituto de Cibern\'etica, Matem\'atica y F\'{\i}sica, Calle
 E 309, Vedado, Ciudad Habana, Cuba}

\begin{abstract}
We theoretically study the coupled modes of a medium-size quantum dot, which
may confine a maximum of ten electron-hole pairs, and a single photonic
mode of an optical microcavity. Ground-state and excitation energies, exciton-photon mixing in the wave functions and the emission of light
from the microcavity are computed as functions of the pair-photon
coupling strength, photon detuning, and polariton number.
\end{abstract}

\maketitle
Semiconductor micropillars  containing  quantum dots in their core region  have been extensively studied recently \cite{Vahala,Yamamoto2002,Gayral} as high efficiency optoelectronic devices \cite{Yariv} and single-photon sources \cite{Imamoglu} due to the enhancement in the spontaneous emission rate (Purcell Effect) \cite{Purcell, Gerard98}. These devices, when operating in the strong coupling regime \cite{Forchel}, could be applied as experimental achievements of quantum information systems, making possible the analysis of entangled photon pairs \cite{Yamamoto2004}, quantum teleportation \cite{Zeilinger}, quantum repeaters \cite{Cirac}, and linear optical quantum computers \cite{Milburn}.

The standard setup is that of a quasi-twodimensional distribution of very small quantum dots coupled to a single cavity mode. The dots are far apart in such a way that we can neglect interactions, and a very simplified model of two-level systems (vacuum - exciton) interacting with the cavity field (the Dicke model \cite{Dicke}) works. The interaction through the photon field forces the polarization functions of isolated dots to be coherent, and makes the physical system a good candidate where BEC of polaritons could be observed \cite{Littlewood}. 

In the present paper, we allow larger dots inside the cavity and study the coupling of a single dot, which may confine a maximum of ten electron-hole pairs, with the photon mode. Exact diagonalization results are presented for ground-state and excitation energies of the coupled system, wave functions, and the energy position and intensity of the light emitted from the cavity as functions of the pair-photon coupling strength and the polariton number. To the best of our knowledge, there are no similar results in the literature.

Although our calculations are intended to be of a qualitative character, we took parameters typical of experimentally studied systems. For example, we think of a GaAs circular micropillar with a diameter of 0.5 $\mu$m, for which the energy separation between the lowest (two-fold degenerated) and first excited cavity modes is around 30 meV. \cite{Gerard} If electronic excitation energies are lower than 30 meV, then the coupling with the excited cavity modes may be neglected. A quality factor of about 20,000 is assumed \cite{Whittaker}, in such a way that, once the cavity is loaded, the photon population decays in around 50 ps. \cite{Pelton} Our calculations for a fixed number of photons (better to say, of polaritons) may be thought off as ``adiabatic'' in nature, corresponding to a given time of the decay process.

A simplified two-band scheme for the GaInAs dots is assumed, with effective in plane masses for the electron and (heavy) hole,  $m_e=0.05~m_0$, and $m_h=0.07~m_0$, respectively. Lateral confinement in the dot is supposed harmonic, with $\hbar\omega_0=4.5$ meV. A magnetic field is impossed along the cavity
axis in order to tune or detune the light from the multiexcitonic levels. For the sake of simplicity, however, in our calculations we fix the magnetic field ($B=3$ Teslas) and move the photon energy around resonance.

The Hamiltonian describing the system is the following:

\begin{eqnarray}
H&=& \sum_{ij}\left\{T^{(e)}_{ij}e^\dagger_i e_j+ T^{(h)}_{\bar i\bar j}h^\dagger_{\bar i} h_{\bar j}\right\} \nonumber\\
&+& \frac{\beta}{2} \sum_{ijkl}\langle ij||kl\rangle~ e^\dagger_i e^\dagger_j e_l e_k +\frac{\beta}{2} \sum_{\bar i\bar j\bar k\bar l}\langle \bar i\bar j||\bar k\bar l\rangle~ h^\dagger_{\bar i} h^\dagger_{\bar j} h_{\bar l} h_{\bar k}
\nonumber\\
&-& \beta \sum_{i\bar j k\bar l}\langle i\bar j||k\bar l\rangle~ e^\dagger_{i} h^\dagger_{\bar j} h_{\bar l} e_{k} + (E_{gap}+\hbar\omega)~ a^\dagger a\nonumber\\
&+& g \sum_i \left\{ a^\dagger h_{\bar i}e_i+a e^\dagger_i h^\dagger_{\bar i}\right\}.
\label{eq1}
\end{eqnarray}

\noindent
The levels $|i\rangle$ correspond to 2D electron states in a magnetic field. The states $|\bar i\rangle$ for holes differ from $|i\rangle$ in the sign of the angular momentum projection along the cavity axis. $T_{ij}=\varepsilon_i\delta_{ij}+\hbar\omega_0^2 \langle i|r^2|j\rangle/\omega_c$, 
where $\varepsilon_i$ is the particle energy in the magnetic field, $\omega_c$ is the cyclotronic frequency, and $r$ is a dimensionless coordinate (coordinates are expressed in units of the magnetic length). $\beta=2.94~\sqrt{B}$ meV, where $B$ is given in Teslas, is the strength of Coulomb interactions, and $\langle ij||kl\rangle$ are the corresponding matrix elements. The energy of the photon mode is written as $E_{gap}+\hbar\omega$, where $E_{gap}$ is the dot effective band gap, and $\hbar\omega$ is a magnitude of the order of a few meV. $g$ is the pair-photon coupling strength. The Hamiltonian (\ref{eq1}) preserves the polariton number:

\begin{equation}
\hat N_{pol}=\hat N_{ph}+\hat N_{pairs},
\end{equation}

\noindent
where $\hat N_{ph}=a^\dagger a$, and $\hat N_{pairs}=\sum_i (h^\dagger_{\bar i}h_{\bar i}+ e^\dagger_i e_i)/2$.

\begin{figure}[t]
\begin{center}
\includegraphics[width=.9\linewidth,angle=0]{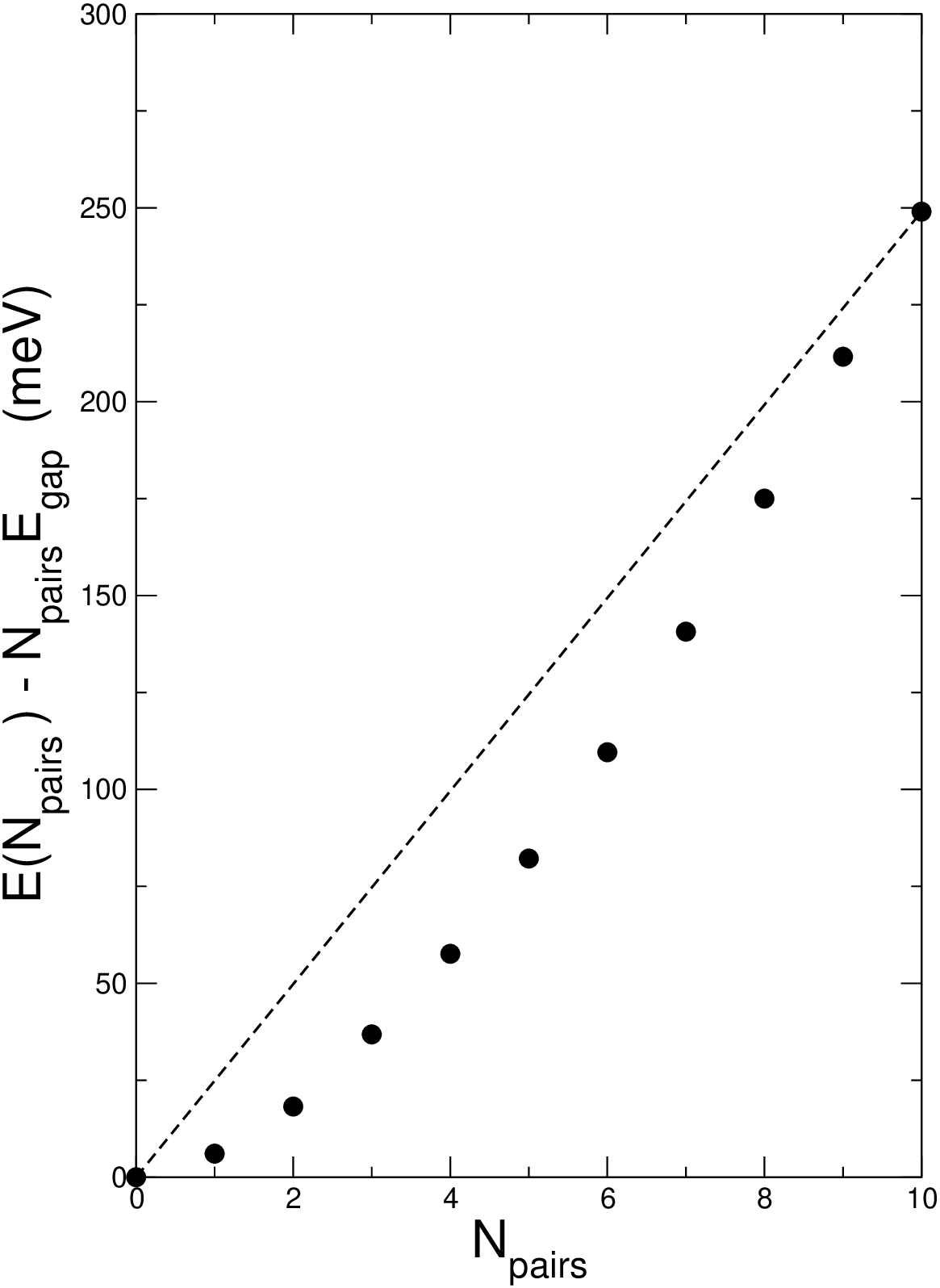}
\caption{\label{fig1} The lowest multiexcitonic levels in the quantum
dot. The reference energy for a state with $N_{pairs}$ is $N_{pairs} E_{gap}$.}
\end{center}
\end{figure}

We show in Fig. \ref{fig1} the lowest multiexcitonic levels (no photons) computed with an energy cutoff of 30 meV in the single-particle orbitals. For a given number of pairs, $N_{pairs}$, a basis set formed from products of Slater determinants for 
electrons and holes, $|S_e\rangle|S_h\rangle$, with total angular momentum projection
equal to zero, is used to diagonalize the electron-hole Hamiltonian.
The reference energy for a level with $N_{pairs}$ is $N_{pairs} E_{gap}$. The dashed line with slope around 25 meV indicates that the resonance condition is $\hbar\omega\approx 25$ meV for our system.

The inclusion of photons is easily understood in the weak coupling regime, $g\to 0$. Both $N_{ph}$ and $N_{pairs}$ are good quantum numbers in this regime. The total energy is:

\begin{eqnarray}
E_{T}&=&N_{ph} (E_{gap}+\hbar\omega) + E(N_{pairs})\nonumber\\
&\approx& N_{pol} (E_{gap}+\hbar\omega) + (25~{\rm meV}-\hbar\omega) N_{pairs}.
\label{eq3}
\end{eqnarray}

\noindent
When $\hbar\omega < 25$ meV (below resonance) the cavity is filled with only photons and no pairs. When $\hbar\omega > 25$ meV (above resonance) the number of pairs in the dot is maximum. And, finally, under resonance conditions, the ground state corresponds to an intermediate occupancy of the dot.

\begin{figure}[t]
\begin{center}
\includegraphics[width=.9\linewidth,angle=0]{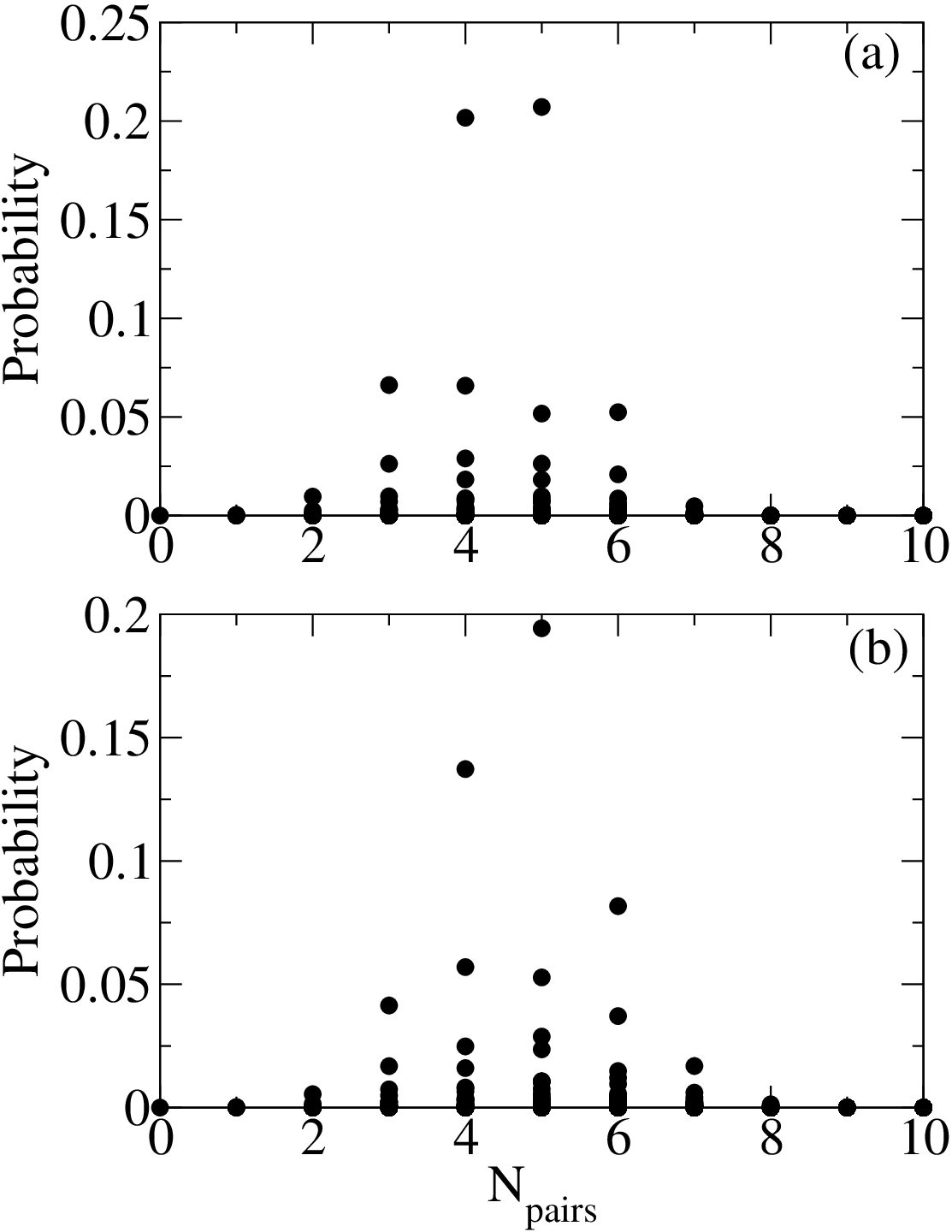}
\caption{\label{fig2} Fock-space probability distributions for the ground-state wavefunctions in cases: (a) $N_{pol}=10$, $g=1$ meV, and (b) $N_{pol}=1000$, $g=0.1$ meV. $\hbar\omega=25$ meV.}
\end{center}
\end{figure}

An increase  of the coupling $g$ increases the mixing between the excitonic and photon modes. An increase of the polariton number has a similar effect, as shown in Fig. \ref{fig2}. The basis set for the exact diagonalization of the Hamiltonian (\ref{eq1})
is now enlarged with the inclusion of the photon component. Basis functions are 
constructed as $|S_e\rangle|S_h\rangle|N_{ph}\rangle$, where $|N_{ph}\rangle$
represents a state with $N_{ph}$ photons. The Fock space probability
distribution for the ground state functions in cases: (a) $N_{pol}=10$, 
$g=1$ meV, and (b) $N_{pol}=1000$, $g=0.1$ meV are schematically drawn in Fig. \ref{fig2}, where the coefficients squared, $|C_{S_e,S_h,N_{ph}}|^2$, are plotted
as functions of $N_{pairs}$, i. e. the number of electron-hole pairs in $(S_e,S_h)$.
Resonance conditions are assumed. 

\begin{figure}[t]
\begin{center}
\includegraphics[width=.9\linewidth,angle=0]{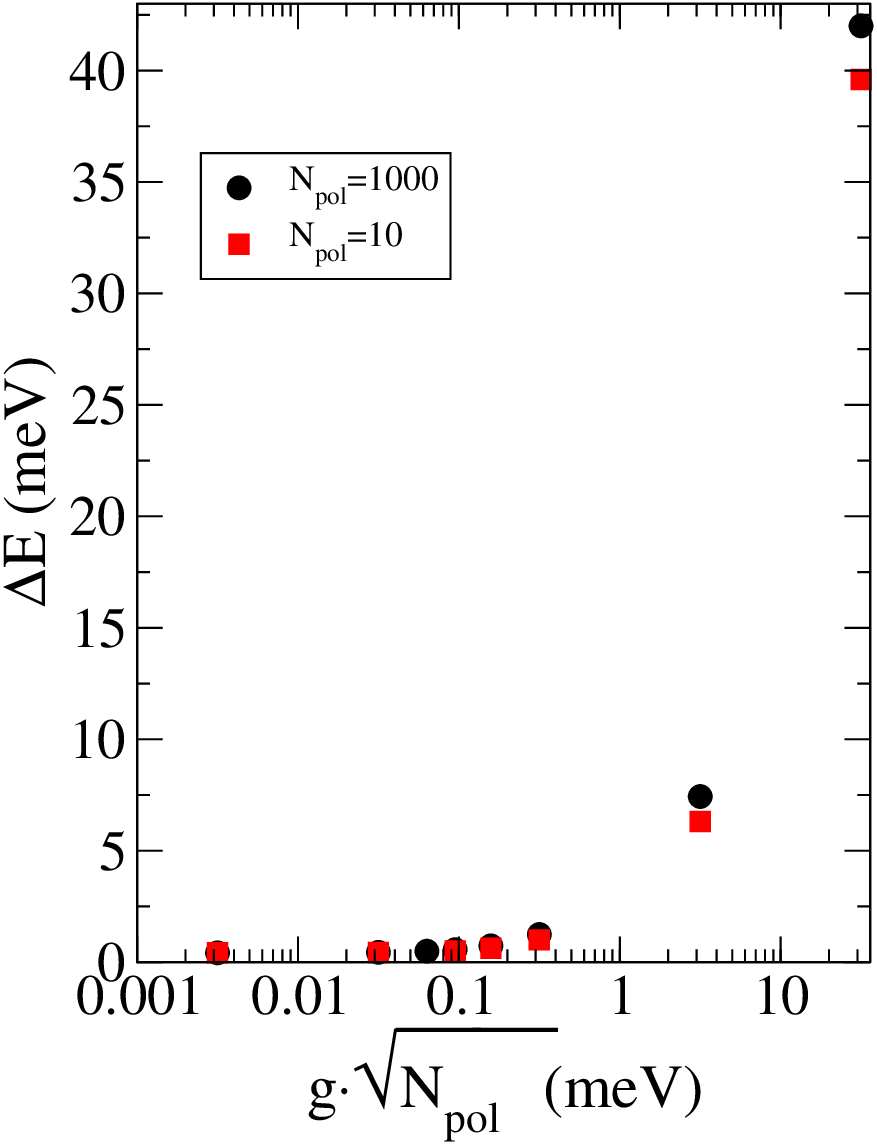}
\caption{\label{fig3} Approximate scaling in the excitation energy. $\hbar\omega=25$ meV.}
\end{center}
\end{figure}

Notice that the highest probabilities correspond to one 5-exciton and one 4-exciton states. Considering only the coupling between the lowest two states in these sectors, we arrive at a simple (qualitative) expression for the excitation energy:

\begin{equation}
\Delta E=\sqrt{A^2+B N_{pol}~g^2},
\label{eq4}
\end{equation}

\begin{figure}[t]
\begin{center}
\includegraphics[width=.9\linewidth,angle=0]{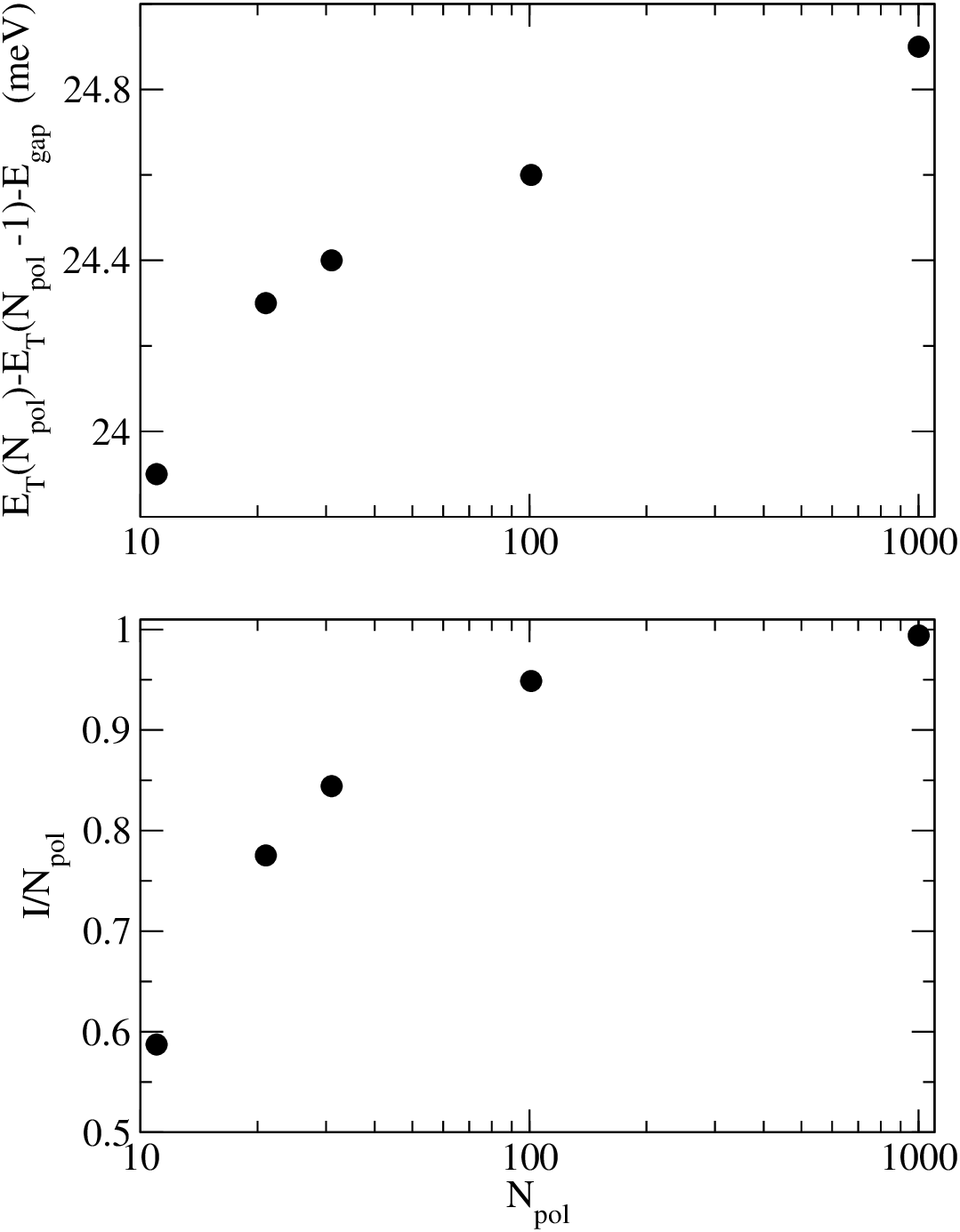}
\caption{\label{fig4} Upper panel: Energy position of the emission line as a function of the polariton number. Lower panel: Relative intensity $I/N_{pol}$ as a function of $N_{pol}$. $\hbar\omega=25$ meV, $g=1$ meV.}
\end{center}
\end{figure}

\noindent
which shows that it is the combination $\sqrt{N_{pol}}~g$ which has the meaning of an effective coupling strength. The constant $A$ in Eq. (\ref{eq4}) has a simple meaning: 
$A=E_{gs}(N_{pairs}=4)+E_{gap}+\hbar\omega-E_{gs}(N_{pairs}=5)$. Approximate scaling in the excitation energy is illustrated in Fig. \ref{fig3} for $N_{pol}=10$ and 1000. $\hbar\omega=25$ meV. We observe an abrupt rise of the excitation gap when $\sqrt{N_{pol}}~g$ exceeds the value 0.1 meV, indicating the emergence of a strongly coupled multiexciton-photon state. Recall that $N_{pol}$ is inversely related to time in the decay process. Then, if the microcavity is initially loaded with a large number of polaritons, the excitation gap will continuosly decrease as time evolves.

Next, we turn to compute the position and intensity of the emission line from the microcavity. The intensity of emitted light is to be obtained from the expression:

\begin{equation}
I \sim |\langle N_{pol}-1|a|N_{pol}\rangle|^2, 
\label{eq5}
\end{equation}

\noindent
where $|N_{pol}\rangle$, $|N_{pol}-1\rangle$ refer to states with  polariton numbers $N_{pol}$ and $N_{pol}-1$, respectively. We assume a working temperature of the order of 1 K, then the initial state is the ground state of the $N_{pol}$ system. Explicit computations from Eq. (\ref{eq5}) show that only the transition to the ground state of the $N_{pol}-1$ system gives a significantly non zero intensity.

We draw in Fig. \ref{fig4} the position and intensity of the emission line as a function of $N_{pol}$. A strong coupling regime, $g=1$ meV, and resonance conditions, $\hbar\omega=25$ meV, are assumed. For large polariton numbers, light is emitted at bare photon energies and $I\approx N_{pol}$, as expected. However, when $N_{pol}$ becomes comparable to the maximum number of pairs in the dot, the emission line is redshifted from $\hbar\omega$, and $I$ significantly deviates from the saturation value, $N_{pol}$.

In conclusion, we presented exact diagonalization results for the coupled states of a multiexcitonic quantum dot and a single photonic mode of a microcavity pillar. Wave function mixing, excitation energies and emission properties were computed as functions of the pair-photon coupling strength, the number of polaritons, and photon detuning. The main result of the paper is the abrupt increase of the excitation gap for $\sqrt{N_{pol}}~g > 0.1$ meV, which opens the possibility for the system to remain in a highly coherent exciton-photon state for a time interval of the order of dozens of picoseconds. Our results could have a relation to the very interesting paper \cite{r18}, where the observation of a polariton system achieving thermal equilibrium with the lattice while decaying is reported. Use of detuning to increase the exciton
components of the wave function, the threshold behavior with the number of polaritons (excitation power) or the almost static position of the emission line (i. e. a shift
of 1 meV when the polariton number is varied by two orders) are all common features.
New calculations in regard to this paper are currently in progress.

The authors are grateful to Hugo Perez Rojas, Alejandro Cabo, and the participants of the Theory Seminar at ICIMAF for useful discussions. The work was partially supported by the Committee for Research of the Universidad de Antioquia and the Centro de Excelencia en Nuevos Materiales, CENM.

\end{document}